\documentclass[final,twoside,11pt]{entics}
\usepackage{quiver}
\usepackage{macros}
\usepackage{graphicx}
\usepackage[all]{xy}
\usepackage{enticsmacro}
\def\conf{MFPS 2023} 	%%Fill in the acronym for your conference (with year)
\def\myconf{mfps39}
\volume{3}			%Fill in the ENTICS volume number here
\def\volu{3}			% and here
\def\papno{20}			%Fill in your paper number here
\def\lastname{van Gool, Melliès, Moreau} %Lastnames appear in the running header 
\def\runtitle{Profinite lambda-terms and parametricity} %Short title appears in the running header on even pages. 
\def\mydoi{12280}
\def\copynames{S. van Gool, P.-A. Melli\'es, V. Moreau}    %% Fill in the first initial and last name of the authors
\begin{document}
%%%Note the beginning and end of the frontmatter section that starts here%%%%%
\begin{frontmatter}
  \title{Profinite Lambda-terms and Parametricity} 						%%Title here and the
  %\thanks[ALL]{General thanks to everyone who should be thanked.}   %%Text of \thanks[ALL} here..
  %%%%%%%%%%%%%%%%%%%%%%%%%%%%			This Thanks is optional.
  %%%%Now the author(s) names(s)%%%%%
  \author{Sam van Gool\thanksref{a}}	%%Note NO SPACE between 
  \author{Paul-André Melliès\thanksref{b}}		%last name and \thanksref{...} 
  \author{Vincent Moreau\thanksref{c}}
  %%%Next come the addresses%%%%
  \address[a]{Université Paris Cité}
  \address[b]{CNRS, Université Paris Cité, Inria}
  \address[c]{Université Paris Cité, Inria}

  \begin{abstract}
    Combining ideas coming from Stone duality and Reynolds parametricity, we formulate in a clean and principled way a notion of profinite $\lambda$-term which, we show, generalizes at every type the traditional notion of profinite word coming from automata theory.
    We start by defining the Stone space of profinite $\lambda$-terms as a projective limit of finite sets of usual $\lambda$-terms, considered modulo a notion of equivalence based on the finite standard model.
    One main contribution of the paper is to establish that, somewhat surprisingly, the resulting notion of profinite $\lambda$-term coming from Stone duality lives in perfect harmony with the principles of Reynolds parametricity.
    In addition, we show that the notion of profinite $\lambda$-term is compositional by constructing a cartesian closed category of profinite $\lambda$-terms,
    and we establish that the embedding from $\lambda$-terms modulo $\beta\eta$-conversion to profinite $\lambda$-terms is faithful using Statman's finite completeness theorem.
    Finally, we prove that the traditional Church encoding of finite words into $\lambda$-terms can be extended to profinite words, and leads to a homeomorphism between the space of profinite words and the space of profinite $\lambda$-terms of the corresponding Church type.
%    \vspace{.75em}
  \end{abstract}
\end{frontmatter}

\section{Introduction}

In this paper, we formulate a notion of \emph{"profinite $\lambda$-term"} which, as we will show, extends in a principled way,
related to Reynolds parametricity,
the important notion of \emph{profinite word} found at the heart of automata theory.

%\medbreak
%0
Our starting point is provided by the Church encoding of finite words
on a given finite alphabet~$\Sigma=\{a_1,\dots,a_n\}$
into simply typed $\lambda$-terms.
The idea of the encoding is to view every letter $a_i\in\Sigma$
as a variable $a_i$ of type $\tyo\To\tyo$ where $\tyo$
is an arbitrary base type.
Once a variable $a_i:\tyo\To\tyo$
has been declared in the context for each letter of~$\Sigma$,
a finite word $w={a_{w_1}} {\cdots} \, {a_{w_k}}\in\Sigma^{\ast}$
can be naturally viewed as the composite
$a_{w_k}\circ \cdots \circ a_{w_1}$ of type $\tyo\To\tyo$.
This composite is represented by the $\lambda$-term $\lambda c. a_{w_k}(\cdots (a_{w_1} c))$, which we note~$W$, where~$c$ is a variable of type~$\tyo$.
The finite word~$w$ %={a_{w_1}} {\cdots} \, {a_{w_k}}\in\Sigma^{\ast}$ 
is thus encoded as the $\lambda$-term defined as
%inline
$\lambda a_1\dots \lambda a_n.\lambda c. a_{w_k}(\cdots (a_{w_1} c))$
which is of type $\Churchs$, defined as
\[
  \underbrace{(\tyo\To\tyo)}_{\mbox{type of $a_1$}}\To\cdots\To \underbrace{(\tyo \To \tyo)}_{\mbox{type of $a_n$}} \To \underbrace{\tyo}_{\mbox{type of $c$}} \To \tyo
  \ ,
\]
where we have $n$ occurrences of $\tyo\To\tyo$,
one for each letter $a_i\in\Sigma$,
and one occurrence of $\tyo$ for the variable~$c$,
on the left of the base type $\tyo$.
\AP Given a simple type~$A$ generated by the base type~$\tyo$, we write $\intro*\LambdaTerms{A}$ for the set of closed $\lambda$-terms of simple type~$A$, considered modulo $\beta$- and $\eta$-conversion.
The Church encoding induces a one-to-one correspondence
\[
  \Sigma^{\ast}
  \  \cong \
  \LambdaTerms{\Churchs}
\]
between finite words on the alphabet~$\Sigma$ and
simply typed $\lambda$-terms of type $\Churchs$ up to $\beta\eta$-equivalence.
The correspondence allows us
to think of finite words on the finite alphabet~$\Sigma$ as
simply typed $\lambda$-terms of that specific type.

\subsubsection*{The finite set interpretation and deterministic automata}
The connection between the Church encoding of finite words
and automata theory has been considered in
syntactic~\cite{hillebrand-kanellakis,salvati,nguyen-pradic-1} and
semantic~\cite{salvati,GrelloisMelliesCSL15,GrelloisMelliesMFPS15,mellies}
contexts.

Here, we follow the semantic track and
focus on the finitary interpretation
of the simply typed $\lambda$-calculus
in the cartesian closed category~$\FinSet$
of finite sets and functions between them,
which, we claim, corresponds to deterministic finite state automata.
\AP In order to define this interpretation,
we start by choosing a finite set~$Q$
which lets us define, for any simple type~$A$,
a finite set $\semset{A}{Q}$ in which we will interpret
$\lambda$-terms of type $A$.
This set $\intro*\semset{A}{Q}$ is inductively defined by
\[
  \semset{\tyo}{Q}\ :=\ Q
  \quad\text{and}\quad
  \semset{A \To B}{Q}\ :=\ \semset{A}{Q} \To \semset{B}{Q}
\]
where we interpret the functional type $A \To B$ as the finite set of set-theoretic functions from the set $\semset{A}{Q}$ to the set $\semset{B}{Q}$.
\AP The interpretation then transports
every simple type $A$
to a finite set $\semset{A}{Q}$ and every simply typed $\lambda$-term~$M$
of type
\[
  a_1:A_1\, , \, \dots \, , \, a_n:A_n \quad \vdash \quad M \, : \, B
\]
to a function between finite sets
\[
  \intro*\semlam{M}{Q}
  \quad : \quad
  \semset{A_1}{Q}\, \times  \, \dots \, \times \, \semset{A_n}{Q}
  \  \longrightarrow \  \semset{B}{Q}
  \ .
\]
This interpretation in $\FinSet$ induces, on closed terms, a function
$\semlam{{-}}{Q} : \LambdaTerms{A} \longrightarrow \semset{A}{Q}$
which is called the ""semantic bracket"" and transports every closed $\lambda$-term~$M$ of type~$A$
to its interpretation
$\semlam{M}{Q} \in \semset{A}{Q}$.
In order to understand the connection with finite automata,
it is instructive to examine how the interpretation acts
on the open $\lambda$-term~$W$
encoding the finite word $w=a_{w_1}\dots a_{w_k}\in\Sigma^*$.
By construction, the $\lambda$-term $W$ is of type
\[
  a_1:\tyo\To\tyo \,, \,\dots \,,\, a_n:\tyo\To\tyo
  \quad \vdash \quad
  W
  \quad : \quad
  \tyo\To\tyo
\]
where each letter~$a_1, \dots, a_n\in\Sigma$ appears as a variable of type $\tyo\To\tyo$ in the context.
The $\lambda$-term~$W$ is then interpreted as the functional
\[
  \semlam{W}{Q}
  \quad:\quad
  (Q\To Q) \times \cdots\times (Q\To Q)
  \ \longrightarrow\
  (Q\To Q)
\]
which transports an $n$-tuple $f_1, \dots, f_n$ of endofunctions
on the finite set~$Q$, i.e. elements of the set $Q \To Q$, to the composite endofunction
$f_{w_k}\circ\dots\circ f_{w_1}$ on the same finite set, that is,
\begin{equation}\label{equation/interpretation-of-W}
  \semlam{W}{Q}
  \  = \
  f_1,\dots,f_n
  \, \mapsto \,
  f_{w_k}\circ\dots\circ f_{w_1}\ .
\end{equation}
A \emph{deterministic finite state automaton} on the alphabet~$\Sigma=\{a_1,\dots,a_n\}$ is defined as a tuple
%inline
$\automaton = (Q, \delta, q_0, \acceptingset)$
consisting of a finite set $Q$ of \emph{states},
a \emph{transition function} $\delta:\Sigma\times Q \to Q$,
an \emph{initial state} $q_0\in Q$
and a set of \emph{accepting states}
$\acceptingset\subseteq Q$.
The transition function~$\delta$ gives rise to a family of transition functions
\[
  \delta_{a_1}=\delta(a_1,-)\,,\,\dots\,,\,
  \delta_{a_n}=\delta(a_n,-)
  \quad : \quad
  Q \ \longrightarrow\ Q
\]
where $\delta_{a}(q)=q'$ means that the automaton~$\automaton$ in state~$q$ transitions to the state~$q'$ when it encounters the letter~${a\in\Sigma}$.

Now observe that, if we apply the interpretation~\eqref{equation/interpretation-of-W} of the simply typed $\lambda$-term~$W$ in $\FinSet$ to these transition functions $\delta_{a_1},\dots, \delta_{a_n}$, then we obtain the endofunction
\[
  \delta_{w} = \semlam{W}{Q}(\delta_{a_1},\dots,\delta_{a_n})
  \quad : \quad
  Q \ \longrightarrow\  Q
\]
which transforms each input state~$q_0\in Q$
into the output state~$q_f=\delta_{w}(q_0)\in Q$
obtained by running the deterministic automaton~$\automaton$
on the finite word~$w$ encoded by the simply typed $\lambda$-term~$W$.
This simple observation establishes the connection between deterministic automata and the interpretation of simply typed $\lambda$-terms of type $\Churchs$ in $\FinSet$.

In this way, if $\automaton = (Q,\delta,q_0,\acceptingset)$ is a deterministic finite state automaton, then the tuple $(Q,\delta,q_0)$
induces an evaluation function
\[
  \evalfun{(\delta,q_0)}
  \quad:\quad
  \semset{\Churchs}{Q}
  \  \longrightarrow \
  Q
\]
which transports every functional $F\in\semset{\Churchs}{Q}$
to the state
%inline
$F(\delta_{a_1},\dots,\delta_{a_k})(q_0)$ in $Q$.
Precomposing the evaluation function $\evalfun{(\delta,q_0)}$ with the "semantic bracket" $\semlam{{-}}{Q}$ induces a composite function
\begin{equation}\label{equation/from-Church-to-states}
  \Sigma^{\ast}\cong\LambdaTerms{\Churchs}
  \  \longrightarrow \
  \semset{\Churchs}{Q}
  \  \longrightarrow \
  Q
\end{equation}
which associates a finite word~$w\in\Sigma^{\ast}$ with the final state $q_f=\delta_{w}(q_0)$ returned by the automaton.
The inverse image of the set~$\acceptingset\subseteq Q$ under
this composite function is, by definition,
the \emph{regular language} $\languageof{\automaton}$ of finite words
recognized by the deterministic automaton~$\automaton$.

\subsubsection*{The Boolean algebra $\reg{A}$ of regular languages}
The regular language~$\languageof{\automaton}$ described above
is an element of the Boolean algebra
$\regset{Q}{\Churchs}$
of regular languages of $\lambda$-terms of type
$\Churchs$ recognized by the finite set $Q$.
This algebra may be defined
as the image of the Boolean algebra homomorphism
$\semlaminv{Q}$
from
$\wp(\semset{\Churchs}{Q})$
to
$\wp(\Sigma^*)$,
obtained by applying the contravariant power set functor
$\wp : {\FinSet}^\mathrm{op} \longrightarrow \BA$
to the "semantic bracket" $\semlam{{-}}{Q}$.
In the theory of regular languages of simply typed $\lambda$-terms developed by Salvati \cite{salvati,SalHdR2015}, this point of view is extended to any type.
The Boolean algebra
$\regset{Q}{A}$
of regular languages of $\lambda$-terms of higher-order type~$A$ recognizable by a finite set~$Q$ of states is defined as the image of the Boolean algebra homomorphism
\[ \semlaminv{Q}
  \quad\colon\quad
  \wp(\semset{A}{Q})
  \hspace{.5em} \longrightarrow \hspace{.5em}
  \wp(\LambdaTerms{A})
  \ .\]
In other words, a set~$L$ of $\lambda$-terms of type~$A$ is recognizable by the finite set~$Q$ precisely when it is of the form
%inline
$\semlaminv{Q}(\acceptingset)
=
\{M\in\LambdaTerms{A} \mid \semlam{M}{Q}\in\acceptingset\}$
for some choice $\acceptingset \subseteq\semset{A}{Q}$ of a set of accepting elements.
Now, letting $Q$ range over all finite sets, the collection
$\reg{A}\ \subseteq\ \wp(\LambdaTerms{A})$
of regular languages of $\lambda$-terms of type~$A$ is defined in~\cite[Def.~1]{salvati} as
\[ \reg{A} \ = \ \bigcup \ \{ \regset{Q}{A} \, \mid \, Q \text{ a finite set}\}.\]
Salvati~\cite[Thm.~8]{salvati} then establishes that $\reg{A}$
is a Boolean algebra, which boils down to the fact that~$\reg{A}$ is closed under intersection. The proof relies on a presentation of higher-order automata based on intersection types, and on the construction of a product higher-order automaton.

\subsubsection*{Profinite words in automata theory}
The monoid $\widehat{\Sigma^\ast}$
of profinite words on a finite alphabet~$\Sigma$ plays
an important role in automata theory,
where profinite words encode the limiting behaviour of
finite words with respect to deterministic finite automata~\cite{pin2009}. For example, one can define an "idempotent power operator" over profinite words, using the fact that the automata are finite, appearing in Equation~\ref{equation/idempotent-power-on-words} within \S\ref{section/proterm-proword} where we explain its construction.
The monoid $\widehat{\Sigma^\ast}$ is the free profinite monoid generated by $\Sigma$ and can be constructed as the limit,
computed in the category $\Mon$ of monoids,
of the codirected
(also known as \emph{projective})
system of finite monoid homomorphisms
\begin{equation}\label{equation/finmon-diagram}\begin{tikzcd}[column sep = 1.2em]
    \Big(\,\,
    \Sigma^\ast/{\phi} \arrow[two heads,rr] && \Sigma^\ast/{\phi'}\Big)_{\phi \subseteq \phi'}
  \end{tikzcd}\end{equation}
where $\phi$ and $\phi'$ range over the finite index congruences on $\Sigma^\ast$, subject to the condition that $\phi \subseteq \phi'$.
Note that every such finite index congruence~$\phi$
can be seen equivalently as a surjective homomorphism
%inline
$h : {\Sigma^\ast} \to M$
to the finite monoid $M = \Sigma^{\ast}/\phi$
whose elements are the equivalence classes of the congruence~$\phi$.
The surjectivity condition on $h$ can be relaxed in order to show that the monoid~$\widehat{\Sigma^\ast}$ of profinite words is in fact the codirected limit of the composite functor
\begin{equation}\label{equation/directed-diagram}
  \begin{tikzcd}[column sep = 1em]
    {\Sigma^{\ast}/\FinMon}
    \arrow[rr,"{\pi}"]
    &&
    {\FinMon}
    \arrow[hook,rr]
    &&
    {\Mon}
  \end{tikzcd}
\end{equation}
where $\FinMon$ denotes the category of finite monoids.
Here, we use the notation ${\Sigma^{\ast}/\FinMon}$
for the slice category whose objects $(M,h)$
are the pairs consisting of a finite monoid~$M$
and of a (not necessarily surjective)
homomorphism of the form
$h:{\Sigma^{\ast}}\to M$,
and whose morphisms $(M,h)\to (M',h')$
are the homomorphisms $f:M\to M'$
making the triangle
\begin{center}
  \begin{tikzcd}[column sep=.8em, row sep = .8em]
    &&
    {\Sigma^{\ast}}
    \arrow[ddll,"{h}"{swap}]
    \arrow[ddrr,"{h'}"]
    \\
    \\
    M\arrow[rrrr,"f"] &&&& M'
  \end{tikzcd}
\end{center}
commute.
The projection functor $\pi$
in \eqref{equation/directed-diagram}
transports $(M,h)$
to the underlying finite monoid~$M$.
One obtains in this way $\widehat{\Sigma^\ast}$
as the limit of a codirected diagram
of finite monoid homomorphisms
\begin{equation}\label{equation/finmon-diagram-bis}\begin{tikzcd}[column sep = 1.2em]
    \Big(\,\,
    M \arrow[rr] && M'\Big)_{(M,h)\to(M',h')}
  \end{tikzcd}\end{equation}
which extends the diagram~\eqref{equation/finmon-diagram} from finite index congruences~$\phi$ on $\Sigma^{\ast}$ to all homomorphisms $h:\Sigma^{\ast}\to M$ to a finite monoid~$M$.

To explain the relationship with automata, recall that every homomorphism $h:{\Sigma^{\ast}}\to M$ to a finite monoid $(M,\cdot_M,e_M)$ induces a deterministic finite automaton, by letting $Q := M$ be the set of states, and defining $\delta(a, q) := q\cdot_M h(a)$ for every letter~$a\in\Sigma$ and state $q\in M$.
This establishes that every monoid homomorphism $h:\Sigma^{\ast}\to M$ to a finite monoid $M=\{q_1,\dots,q_m\}$
induces a decomposition of~$\Sigma^{\ast}$
into $m$ components
$\languageof{q_i} = h^{-1}(q_i)$
for $1\leq i\leq m$
where each
$\languageof{q_i}$ is a regular language.
We will denote by $\intro*\regmon{(M,h)}{\Sigma}$ the Boolean algebra of languages generated by the regular languages of the form $\languageof{q}=h^{-1}(q)$, as $q$ ranges over the elements of $M$.
One obtains in this way a functor
\begin{equation}\label{equation/contravariant-functor}
    \regmon{(-)}{\Sigma}
    \quad:\quad
    {(\Sigma^{\ast}/\FinMon)}^\mathrm{op}
    \ \longrightarrow\ 
    {\BA}
\end{equation}
to the category $\BA$ of Boolean algebras,
which maps every pair $(M,h)$ to
the Boolean algebra $\regmon{(M,h)}{\Sigma}$.
Note that this Boolean algebra $\regmon{(M,h)}{\Sigma}$
coincides with the image of the Boolean algebra homomorphism
$h^{-1} : \wp(M) \longrightarrow \wp(\Sigma^*)$
obtained by applying the contravariant powerset functor $\wp$
to the map $h \colon \Sigma^* \to M$.
An important insight of \cite[Sec.~4.2]{Geh2016} is that
the following directed diagram in $\BA$, associated to the functor~\eqref{equation/contravariant-functor},
\begin{center}\begin{tikzcd}[column sep = 1.2em]
  \Big(\,\,
  \regmon{(M',h')}{\Sigma} \arrow[hook,rr] && \regmon{(M,h)}{\Sigma}
  \Big)_{(M, h) \to (M', h')}
\end{tikzcd}\end{center}
may be obtained more directly by applying~$\wp$
to the codirected diagram of finite sets
underlying~\eqref{equation/directed-diagram}
and~\eqref{equation/finmon-diagram-bis}.
Since the colimit in $\BA$
of this diagram coincides with $\operatorname{Reg}(\Sigma)$,
one establishes in this way
that the monoid of profinite words
is in fact the Stone dual of the Boolean algebra~$\operatorname{Reg}(\Sigma)$ of regular sets,
see \cite{Geh2016} as well as \S\ref{section/regular} below
for details.

\subsubsection*{From profinite words to profinite $\lambda$-terms}
In order to define the notion of
\emph{"profinite $\lambda$-term"}
at an arbitrary simple type $A$,
we combine this general scheme
with ideas coming from Reynolds parametricity.
We have seen that, given a finite set $Q$, we can interpret any simple type $A$ as a finite set $\semset{A}{Q}$.
To relate elements belonging to two different interpretations
$\semset{A}{Q}$ and $\semset{A}{Q'}$
one can construct, given a relation
$R \subseteq Q \times Q'$ between the finite sets
used for the interpretation, a relation
$\semrel{A}{R}\ \subseteq\ \semrel{A}{Q} \times \semrel{A}{Q'}$
between the two interpretations of the simple type~$A$.
Such inductively-defined relations are called
\emph{"logical relations"}.
A fundamental fact is that~$\lambda$-terms are parametric,
that is, for any $\lambda$-term~$M$ of type~$A$ and any
relation~$R \subseteq Q \times Q'$,
\[
  (\semlam{M}{Q}, \semlam{M}{Q'})\ \in\ \semrel{A}{R}\ .
\]

In particular, we will recall in
Proposition~\ref{prop/partial-surjections}
the well-known fact that
every "partial surjection"~$f~:~Q~\twoheadrightarrow~Q'$,
seen as a relation,
induces a "partial surjection"
$\semrel{A}{f} : \semset{A}{Q} \twoheadrightarrow \semset{A}{Q'}$
such that for every $\lambda$-term~$M$ of type~$A$,
its interpretation $\semlam{M}{Q}$
is in the domain of $\semrel{A}{f}$
and the "partial surjection" $\semrel{A}{f}$ sends
the interpretation of~$M$
in the finite set $\semset{A}{Q}$ to its interpretation
in $\semset{A}{Q'}$,
that is,
\begin{equation}\label{equation/terms-parametric}
  \semrel{A}{f}\left(\semlam{M}{Q}\right) \ = \ \semlam{M}{Q'}\ .
\end{equation}
An easy argument, given in
Lemma~\ref{lemma/partial-surj-inclusion} below, shows that,
as a consequence, every "partial surjection"
$f:Q\twoheadrightarrow Q'$ induces
an inclusion of Boolean algebras $\regset{Q'}{A} \subseteq \regset{Q}{A}$.
We note $\FinPSurj$ the category whose objects are finite sets and whose morphisms are "partial surjections". For every simple type~$A$, we then have a functor
\[
  \regset{(-)}{A}
  \quad:\quad
  \FinPSurj^{\mathrm{op}}
  \ \longrightarrow\ 
  \BA
\]
which sends each "partial surjection" on the associated inclusion of Boolean algebras.
This leads us to the first main result
of the paper, established in \S\ref{section/regular}.
\medbreak
\noindent
\AP\textbf{\emph{""Theorem~A"".}}
\emph{The diagram of Boolean algebras $\regset{(-)}{A} : \FinPSurj^{\mathrm{op}} \to \BA$, i.e.}
\begin{center}\begin{tikzcd}[column sep = 1.2em]
    \Big(\,\, %\label{equation/regA-diagram}
    \regset{Q'}{A}  \arrow[hook,rr] && \regset{Q}{A}
    \Big)_{f:Q\twoheadrightarrow Q'\in\FinPSurj}
    \quad,
  \end{tikzcd}\end{center}
\emph{is directed, and its colimit
  in $\BA$ coincides with the Boolean
  algebra~$\reg{A}$ of regular languages
  of higher-order type~$A$.}

\medbreak

\noindent
\AP
At this stage, a key observation coming from Stone duality is that, for each finite set $Q$, the finite Boolean algebra $\regset{Q}{A}$ is join-generated by its finite set of atoms,
which, as we will show in Proposition~\ref{prop/atomsofreg} below, is in bijection with the set
\[
  \intro*\definable{A}{Q}
  \ =\
  \Big\{\, \semlam{M}{Q} \mid M\in\LambdaTerms{A} \,\Big\}
  \ \subseteq \ \semset{A}{Q}
\]
of ""definable elements"" in $\semset{A}{Q}$.
Moreover,
using (\ref{equation/terms-parametric}),
we see that,
for every "partial surjection" $f \colon Q \twoheadrightarrow Q'$,
there exists a unique (total) surjection
%inline
$\definable{A}{f} : \definable{A}{Q} \twoheadrightarrow \definable{A}{Q'}$
making the following diagram commute
\begin{center}
  \begin{tikzcd}[column sep = 3em, row sep = 1em]
    \definable{A}{Q}
    \arrow[rr,two heads,"{\definable{A}{f}}"]
    \arrow[dd,hook]
    && \definable{A}{Q'}
    \arrow[dd,hook]
    \\
    \\
    \semset{A}{Q} \arrow[rr,two heads,"{\semrel{A}{f}}"]
    &&
    \semset{A}{Q'}
  \end{tikzcd}
\end{center}
in the category $\FinPSet$ of finite sets and partial functions.
We are now ready to define the set $\ProLambdaTerms{A}$
of "profinite $\lambda$-terms" of type~$A$
as the limit in the category~$\Set$ of the codirected diagram
of finite sets
\begin{center}\begin{tikzcd}[column sep = 1.2em]
    \Big(\,\,
    \definable{A}{f}
    \  : \
    \definable{A}{Q} \arrow[two heads,rr] && \definable{A}{Q'}\Big)_{f:Q\twoheadrightarrow Q'\in\FinPSurj}
  \end{tikzcd}\end{center}
indexed by "partial surjections" between finite sets. This diagram is dual
to the directed diagram in $\BA$
defining the Boolean algebra $\reg{A}$ in "Theorem~A".
Moreover, by Stone duality, the set~$\ProLambdaTerms{A}$ of "profinite $\lambda$-terms"
of type~$A$ is not just a set, but a Stone space,
dual to the Boolean algebra~$\reg{A}$.

The conceptual definition of "profinite $\lambda$-term" which we have just given is nice but probably a little bit abstract to a reader with expertise in the $\lambda$-calculus but not necessarily in Stone duality.
A more pedestrian way to understand it is to think of a "profinite $\lambda$-term"~$\theta \in \ProLambdaTerms{A}$ of type~$A$ as 
a family of "definable elements"
$\theta_Q \  \in \  \definable{A}{Q}$
indexed by finite sets $Q$,
such that the family $\theta$
is moreover \emph{natural}
{with respect to finite "partial surjections"},
in the expected sense that the equality $\definable{A}{f} (\theta_Q) \ = \ \theta_{Q'}$
holds for every "partial surjection"~$f : Q\twoheadrightarrow Q'$
between finite sets.

\subsubsection*{Profinite $\lambda$-terms and Reynolds parametricity}
The pedestrian definition of "profinite $\lambda$-terms" just given requires
that the family of definable elements $\theta_Q\in\definable{A}{Q}$
is natural with respect to finite "partial surjections", instead of asking
the stronger property that the family~$\theta$
is \emph{parametric} in the traditional sense of Reynolds.
We establish in \S\ref{section/profiniteness-and-parametricity} the important property
that every "profinite $\lambda$-term" may be equivalently defined using parametricity instead of partial surjections, as follows:

\medbreak
\noindent\AP\textbf{\emph{""Theorem~B"".}}
\emph{
  A "profinite $\lambda$-term" $\theta \in \ProLambdaTerms{A}$
  of type~$A$
  may be equivalently defined as a family of 
  definable elements $\theta_Q\in\definable{A}{Q}$
indexed by finite sets $Q$,
such that the family $\theta$
is moreover \emph{parametric}
with respect to any "logical relation",
in the sense that
$(\theta_Q,\theta_{Q'})\in\semrel{A}{R}$
for every relation $R \subseteq Q \times Q'$.}
\medbreak
\noindent
As we will see in \S\ref{section/profiniteness-and-parametricity},
the fact that the notion based on parametricity
is stronger than the notion based on naturality
is easy to show. What is more difficult to establish
that the two notions are in fact equivalent.

\subsubsection*{The cartesian closed category $\ProLam$ of profinite lambda-terms}
We establish that the resulting
notion of "profinite $\lambda$-term" is compositional
by constructing a cartesian closed category $\ProLam$
of "profinite $\lambda$-terms".
There is a functor
\[
\inclusion{}{}
\quad:\quad
\Lam
\ \longrightarrow\ 
\ProLam
\]
which is faithful by Statman's theorem and which embeds the simply typed $\lambda$-terms into "profinite $\lambda$-terms". It associates to a simply typed $\lambda$-term $M$ the "profinite $\lambda$-term" whose component at the finite set $Q$ is the interpretation $\semlam{M}{Q}$.

Another interesting fact is that there exists, for every simple type~$A$,
a "profinite $\lambda$-term" defining a fixpoint operator
\begin{center}
  $\Omega_A \ \in \ \ProLambdaTerms{(A\Rightarrow A)\Rightarrow (A\Rightarrow A)}$
\end{center}
which thus defines a morphism
\begin{center}
  \begin{tikzcd}[column sep = 1em]
    \Omega_A \quad : \quad (A\Rightarrow A)
    \arrow[rr] && (A\Rightarrow A)
  \end{tikzcd}
\end{center}
in the category~$\ProLam$ of "profinite $\lambda$-terms".
The fixpoint operator~$\Omega_A$ is similar in spirit but different in practice from the usual fixpoint operators $Y_A:(A\Rightarrow A)\Rightarrow A$ of Scott domain semantics, and one interesting direction for future work will be to understand how the two fixpoint operators $\Omega_A$ and $Y_A$ are related.

We also establish at the end of the paper (see \S\ref{section/proterm-proword})
that "profinite $\lambda$-terms" of type $\Churchs$
are the same thing as profinite words over the alphabet $\Sigma$ in the traditional sense.
\medbreak
\noindent
\AP\textbf{\emph{""Theorem~C"".}}
\emph{For every finite set $\Sigma$, there is a homeomorphism between the space of "profinite $\lambda$-terms" of type~$\Churchs$ and the space of profinite words over $\Sigma$, that is,}
\[
  \ProLambdaTerms{\Churchs}\ \cong\ \widehat{\Sigma^*}
  \ .
\]

\subsubsection*{Related works}
As explained in the introduction,
our present definition of "profinite $\lambda$-term"
relies on the notion of \emph{regular language}
of simply typed $\lambda$-terms
introduced by Salvati~\cite{salvati}.
Interestingly, the notion of regular language
is formulated by Salvati
in two different but equivalent ways.
The first definition of regular language
is based on the interpretation of $\lambda$-terms
in the finite standard model $\FinSet$
of the simply typed $\lambda$-calculus.
This is the definition which we recall
and develop in the introduction and in the paper.
The second equivalent definition given by Salvati
relies on the construction
of an intersection type system in direct correspondence
with the finite monotone model of the simply typed $\lambda$-calculus constructed in the category~$\FinScott$ of finite lattices and monotone maps between them,
see~\cite{SalHdR2015} for a discussion.
Aware of this correspondence with Scott semantics, Salvati and Walukiewicz actively promoted a semantic approach to higher-order model checking \cite{salvati-walukiewicz} which would complement
the intersection type approach developed by Kobayashi and Ong \cite{kobayashi,kobayashi-ong}.
However, besides the fascinating connections to Krivine environment machines
and collapsible pushdown automata \cite{DBLP:journals/iandc/SalvatiW14,HagueMOS17,BroadbentCOS21},
it took several years to develop a precise connection between Scott semantics and intersection type systems for higher-order model checking, with
the emergence of a notion of higher-order parity automaton~\cite{mellies}
founded on the discovery of an unexpected relationship with linear logic~\cite{bucciarelli-ehrhard-1,bucciarelli-ehrhard-2,GrelloisMelliesCSL15,GrelloisMelliesMFPS15} combined with a comonadic translation designed by Melliès
of the simply typed $\lambda Y$-calculus into a $\lambda Y_{\mu\nu}$-calculus with inductive and coinductive fixpoints~\cite{mellies}, or into a $\lambda Y$-calculus with priorities~\cite{Walukiewicz19}.

One fundamental idea which emerged from these works,
also apparent in the work by Colcombet and Petrişan~\cite{colcombet-petrisan},
is that there exists a correspondence
between the specific category used
for the semantic interpretation
and a specific class of automata of interest.
Typically, the interpretation of the simply typed $\lambda$-calculus in $\FinSet$ corresponds
to the class of deterministic automata, while the interpretation in $\FinScott$
corresponds to the class of non-deterministic automata.
In the present paper, we focus on the finite standard model in $\FinSet$,
and leave the investigation of the finite monotone lattice model in $\Scott$ for future works.

Another important line of work at the interface of automata theory
and $\lambda$-calculus was initiated by Hillebrand and Kanellakis~\cite{hillebrand-kanellakis} with a {purely syntactic} description of regular languages of finite words using the Church encoding in the simply typed $\lambda$-calculus.
This alternative approach is extremely promising and has seen a recent revival with
the works by Nguy{\^{e}}n and Pradic on implicit automata theory~\cite{nguyen-pradic-1,nguyen-pradic-2}.
Our definition of "profinite $\lambda$-term" is formulated using the finite standard model, but it is largely independent of it, and it would thus be interesting to recast our definition of "profinite $\lambda$-term"
in this purely syntactic framework.

In the study of regular languages and
profinite monoids, the potential role of Stone duality
was identified early on by Pippenger~\cite{Pip1997}, and can also already
be recognized in the ``implicit operations'' which were introduced by
Reiterman~\cite{reiterman} and play a role in
Almeida's important work on profinite semigroups~\cite{almeida2005}.
It is also interesting to note in this context
that monoidal relations, under the name
of ``relational morphisms'', have long played an important role in
(pro)finite semigroup theory, as exemplified for example
by Rhodes and Steinberg~\cite{RS2008},
and our crucial use of logical relations in this paper
opens up potential new connections with that theory.

The specific methodology of understanding
profinite algebraic structure by applying
Stone duality to a lattice of regular languages
that we closely follow in
Sections~\ref{section/regular}~and~\ref{section/profinite} of
this paper emerged from an
influential series of works by Gehrke, Grigorieff and Pin~\cite{GGP2008,GGP2010},
culminating in Gehrke's~\cite{Geh2016}, which contains
the most general account to date of that line of research.
In a direction that is related to, but different from, the one
pursued in this paper, Boja{\'n}czyk~\cite{Boj15b}
generalized these profinite ideas
to the category of algebras given by an arbitrary monad,
also see the more recent work by Ad{\'a}mek et al.~\cite{ACMU21}
pursuing a similar direction.
While these works were always based in an algebraic
setting, a novel contribution of this paper
is to show how these ideas extend to the
setting of the simply typed $\lambda$-calculus and cartesian closed categories.

\subsubsection*{Overview of the paper}
We start by recalling in \S\ref{section/regular}
the notion of regular language of $\lambda$-terms induced by the finite standard model of the simply typed $\lambda$-calculus.
Then, as explained in the introduction, we establish in \S\ref{section/regular} that the Boolean algebra $\reg{A}$ of regular languages of simply typed $\lambda$-terms of type $A$ formulated by Salvati can be equivalently expressed ("Theorem~A") as a colimit in $\BA$ of a specific directed diagram of finite Boolean algebras $\regset{Q}{A}$.
This leads us to introduce in~\S\ref{section/profinite}
the set $\ProLambdaTerms{A}$ of \emph{"profinite $\lambda$-terms"} of type~$A$, which we define as the limit in $\Set$ of a specific codirected diagram of finite sets $\definable{A}{Q}\subseteq \semset{A}{Q}$.
We also show that, by construction, the set %$\ProLambdaTerms{A}$ 
of "profinite $\lambda$-terms"
can be equipped with a natural topology which turns $\ProLambdaTerms{A}$ into the
Stone space
dual to the Boolean algebra~$\regset{Q}{A}$.
We establish in the next section \S\ref{section/profiniteness-and-parametricity}
that "profinite $\lambda$-terms" can be defined
in an alternative and more direct way as
families of "definable elements" $\theta_Q\in\definable{A}{Q}$ satisfying a parametricity property with respect
to any binary relation $R\subseteq Q\times Q'$.
This is the essence of "Theorem~B" mentioned
in the introduction.
We then show in \S\ref{section/profinite-ccc}
that the resulting notion of "profinite $\lambda$-term" is compositional in the technical sense that it defines a cartesian closed category~$\ProLam$
whose objects are the simply types
and whose morphisms are "profinite $\lambda$-terms".
Using Statman's theorem, we establish in \S\ref{section/faithful-embedding}
that the canonical functor from the category~$\Lam$ of simply typed $\lambda$-terms to the category~$\ProLam$
of "profinite $\lambda$-terms" is a faithful embedding.
Finally, we establish in \S\ref{section/proterm-proword}
our theorem ("Theorem~C") that given a finite alphabet~$\Sigma$ of letters, the notion of "profinite $\lambda$-terms" of type $\Churchs$ coincides with the usual notion of profinite words over $\Sigma$.
We conclude and give a number of perspectives for future work in~\S\ref{section/conclusion}.

\section{Regular languages of \texorpdfstring{$\lambda$}{λ}-terms}\label{section/regular}
In this section, we define the collection $\reg{A}$ of
"regular languages" at an arbitrary type $A$,
and establish "Theorem~A" of the introduction,
showing how $\reg{A}$ can be built as the colimit of a
directed diagram in the category of Boolean algebras.

\begin{definition}
  \AP
  Let $Q$ be a finite set and $A$ a type. We say that a
  subset $L \subseteq \LambdaTerms{A}$ is a ""regular language""
  of type $A$ recognized by $Q$ if
  there exists a subset $\acceptingset$ of $\semset{A}{Q}$ such
  that $L = \semlaminv{Q}(\acceptingset)$, that is,
  \[
    \text{for any } M \in \LambdaTerms{A}, \quad M \in L \iff \semlam{M}{Q} \in \acceptingset \ .
  \]
  We denote the Boolean algebra of "regular languages" of
  type $A$ recognized by $Q$ by
  %inline
  $\intro*\regset{Q}{A} \subseteq \wp(\LambdaTerms{A})$
  and we write
  \[\intro*\reg{A} \ =\ \bigcup\, \left\{ \regset{Q}{A} \mid Q \text{ a finite set} \right\} \ \]
  for the collection of "regular languages" of type $A$.
\end{definition}
While it is clear that, for each individual
finite set $Q$, the set $\regset{Q}{A}$ is closed under
the Boolean operations,
since it is defined as the image of the Boolean homomorphism
$\semlaminv{Q}$,
it is not immediately apparent that the union $\reg{A}$
is also closed
under the Boolean operations.

\AP To this end, we will use ""logical relations"". If $S \subseteq P \times P'$ and $R \subseteq Q \times Q'$ are two set-theoretic relations between finite sets, then one can define their exponential, which is $S \To R \subseteq (P \To Q) \times (P' \To Q')$, as
\[
  S \To R
  \quad:=\quad
  \{(g, h) \mid \text{for all $x\ R\ y$, we have $g(x)\ R'\ h(y)$}\}
  \ .
\]
\AP Therefore, for any relation $R \subseteq Q \times Q'$ between two finite sets $Q$ and $Q'$, we construct the relation $\intro*\semrel{A}{R} \subseteq \semset{A}{Q} \times \semset{A}{Q'}$ by induction on the simple type $A$ as
\[
  \semrel{\tyo}{R}
  \ :=\ 
  R
  \qquad\text{and}\qquad
  \semrel{A \To B}{R}
  \ :=\ 
  \semset{A}{R} \To \semset{B}{R}
  \ .
\]
\AP The ""fundamental lemma of logical relations"" then states that for all $M \in \LambdaTerms{A}$ and any $R \subseteq Q \times Q'$, the interpretations of $M$ at $Q$ and $Q'$ are related in the sense that
\[
  \semlam{M}{Q}
  \quad
  \semrel{A}{R}
  \quad
  \semlam{M}{Q'}
  \ .
\]
\AP In particular, we will make extensive use of ""partial surjections"", i.e. relations wich are graphs of surjective partial functions. We note $f : Q \twoheadrightarrow Q'$ such a "partial surjection". We first prove the following lemma, which states that "partial surjections" are stable by exponential.

\begin{lemma}\label{lemma/arrow-part-surj}
  If $e : P \twoheadrightarrow P'$ and $f : Q \twoheadrightarrow Q'$ are "partial surjections", then so is the relation $e \To f$.
\end{lemma}

\begin{proof}
  We first remark that a "partial surjection" $f : Q \twoheadrightarrow Q'$ may equivalently be described as a span
  \[
    % file:///Users/vimo/Documents/Latex/quiver/src/index.html?q=WzAsMyxbMSwwLCJmIl0sWzAsMCwiUSJdLFsyLDAsIlEnIl0sWzAsMiwiXFxwaV8yIl0sWzAsMSwiXFxwaV8xIiwyXV0=
\begin{tikzcd}[ampersand replacement=\&]
	Q \& f \& {Q'}
	\arrow["{\pi_2}", from=1-2, to=1-3]
	\arrow["{\pi_1}"', from=1-2, to=1-1]
\end{tikzcd}
  \]
  where $\pi_1$ is injective and $\pi_2$ is surjective. We adopt this viewpoint during this proof.

  Let~$e : P \twoheadrightarrow P'$ and~$f : Q \twoheadrightarrow Q'$ be two "partial surjections". Note that two functions~$g : P \to Q$ and~$h : P' \to Q'$ are related by~$e \To f$ if and only if there exists a function~$c : e \to f$ such that the following diagram commutes:
  \begin{equation}\label{equation/cgh}
% file:///Users/vimo/Documents/Latex/quiver/src/index.html?q=WzAsNixbMSwwLCJlIl0sWzAsMCwiUCJdLFsyLDAsIlAnIl0sWzEsMSwiZiJdLFswLDEsIlEiXSxbMiwxLCJRJyJdLFswLDEsIlxccGlfMSIsMl0sWzAsMiwiXFxwaV8yIl0sWzAsMywiYyJdLFszLDUsIlxccGlfMiciXSxbMyw0LCJcXHBpXzEnIiwyXSxbMiw1LCJoIl0sWzEsNCwiZyIsMl1d
\begin{tikzcd}[ampersand replacement=\&]
	P \& e \& {P'} \\
	Q \& f \& {Q'}
	\arrow["{\pi_1}"', from=1-2, to=1-1]
	\arrow["{\pi_2}", from=1-2, to=1-3]
	\arrow["c", from=1-2, to=2-2]
	\arrow["{\pi_2'}", from=2-2, to=2-3]
	\arrow["{\pi_1'}"', from=2-2, to=2-1]
	\arrow["h", from=1-3, to=2-3]
	\arrow["g"', from=1-1, to=2-1]
\end{tikzcd}
    \ .
  \end{equation}

  First, we show that the relation $e \To f$ is a partial function. Let $g : P \to Q$ and $h_1, h_2 : P' \to Q'$ such that $(g, h_i) \in e \To f$ for~$i = 1, 2$.
  We thus have two maps~$c_i : e \to f$ for~$i = 1, 2$ such that
  \[
    g \circ \pi_1 = \pi_1' \circ c_i
    \quad\text{and}\quad
    h_i \circ \pi_2 = \pi_2' \circ c_i
    \quad\text{for $i = 1, 2$}
    \ .
  \]
  The maps all fit in the diagram
  \[
% file:///Users/vimo/Documents/Latex/quiver/src/index.html?q=WzAsNixbMSwwLCJlIl0sWzAsMCwiUCJdLFsyLDAsIlAnIl0sWzEsMSwiZiJdLFswLDEsIlEiXSxbMiwxLCJRJyJdLFswLDEsIlxccGlfMSIsMl0sWzAsMiwiXFxwaV8yIl0sWzAsMywiY18yIiwwLHsiY3VydmUiOi0xfV0sWzMsNSwiXFxwaV8yJyJdLFszLDQsIlxccGlfMSciLDJdLFsyLDUsImhfMiIsMCx7ImN1cnZlIjotMX1dLFsxLDQsImciLDJdLFswLDMsImNfMSIsMix7ImN1cnZlIjoxfV0sWzIsNSwiaF8xIiwyLHsiY3VydmUiOjF9XV0=
\begin{tikzcd}[ampersand replacement=\&]
	P \& e \& {P'} \\
	Q \& f \& {Q'}
	\arrow["{\pi_1}"', from=1-2, to=1-1]
	\arrow["{\pi_2}", from=1-2, to=1-3]
	\arrow["{c_2}", curve={height=-6pt}, from=1-2, to=2-2]
	\arrow["{\pi_2'}", from=2-2, to=2-3]
	\arrow["{\pi_1'}"', from=2-2, to=2-1]
	\arrow["{h_2}", curve={height=-6pt}, from=1-3, to=2-3]
	\arrow["g"', from=1-1, to=2-1]
	\arrow["{c_1}"', curve={height=6pt}, from=1-2, to=2-2]
	\arrow["{h_1}"', curve={height=6pt}, from=1-3, to=2-3]
\end{tikzcd}
    \ .
  \]
  By injectivity of $\pi_1'$, we get that $c_1 = c_2$. Therefore,
  \[
    h_1 \circ \pi_2 = \pi_2' \circ c_1 = \pi_2' \circ c_2 = h_2 \circ \pi_2
    \ .
  \]
  By surjectivity of $\pi_2$, we get that $h_1 = h_2$. This proves that the relation $e \To f$ is a partial function.

  We now show that the relation~$e \To f$ is surjective. Let~$h \colon P' \to Q'$ be any function.
  As~$\pi_2'$ is surjective, it has a section~$s \colon Q' \to f$, that is, $\pi_2' \circ s = \mathrm{id}_{Q'}$. We then define the function~$c \colon e \to f$ as
  $s \circ h \circ \pi_2$.
  Note that
  $\pi_2' \circ c = h \circ \pi_2$,
  so we obtain the commuting diagram
  \[
% file:///Users/vimo/Documents/Latex/quiver/src/index.html?q=WzAsNixbMSwwLCJlIl0sWzAsMCwiUCJdLFsyLDAsIlAnIl0sWzEsMSwiZiJdLFswLDEsIlEiXSxbMiwxLCJRJyJdLFswLDEsIlxccGlfMSIsMl0sWzAsMiwiXFxwaV8yIl0sWzAsMywiYyJdLFszLDUsIlxccGlfMiciXSxbMyw0LCJcXHBpXzEnIiwyXSxbMiw1LCJoIl0sWzUsMywicyIsMCx7ImN1cnZlIjotMX1dXQ==
\begin{tikzcd}[ampersand replacement=\&]
	P \& e \& {P'} \\
	Q \& f \& {Q'}
	\arrow["{\pi_1}"', from=1-2, to=1-1]
	\arrow["{\pi_2}", from=1-2, to=1-3]
	\arrow["c", from=1-2, to=2-2]
	\arrow["{\pi_2'}", from=2-2, to=2-3]
	\arrow["{\pi_1'}"', from=2-2, to=2-1]
	\arrow["h", from=1-3, to=2-3]
	\arrow["s", curve={height=-6pt}, from=2-3, to=2-2]
\end{tikzcd}
    \ .
  \]
  As $\pi_1$ is injective, it has a retraction $r : P \to e$, that is, $r \circ \pi_1 = \mathrm{id}_{e}$. We define the function~$g : Q \to Q'$ as
  $\pi_1' \circ c \circ r$
  and note that
  \[
    g \circ \pi_1
    \ =\
    \pi_1' \circ c
    \ .
  \]
  Thus, the diagram (\ref{equation/cgh}) commutes for this choice of $g$, $c$, and $h$, which means that $(g, h)$ is in the relation $e \To f$, as required. This shows that $e \To f$ is surjective.
\end{proof}

Using Proposition~\ref{lemma/arrow-part-surj}, we get a proof of the following proposition by induction on simple types.

\begin{prop}\label{prop/partial-surjections}
  If $f \colon Q \twoheadrightarrow Q'$ is a "partial surjection", then the relation $\semrel{A}{f} \subseteq \semset{A}{Q} \times \semset{A}{Q'}$ is a "partial surjection" for any simple type $A$.
\end{prop}

The following lemma contains the crucial argument
needed to prove the fact that $\reg{A}$ is a Boolean algebra.
\begin{lemma}\label{lemma/partial-surj-inclusion}
  Let $f \colon Q \twoheadrightarrow Q'$ be a
  "partial surjection".
  Then, for any simple type $A$, we have an inclusion
  of Boolean algebras $\regset{Q'}{A} \subseteq \regset{Q}{A}$.
\end{lemma}
\begin{proof}
  Let $A$ be any simple type and let $\acceptingset' \subseteq \semset{A}{Q'}$,
  recognizing the language
  $L = \semlaminv{Q'}(\acceptingset')$
  in $\regset{Q'}{A}$.
  We define the subset $\acceptingset$ of $\semset{A}{Q}$ as $\semrelinv{A}{f}(\acceptingset')$, that is
  $\{ x \in \semset{A}{Q} \mid x \, \semrel{A}{f} \, y \text{ for some } y \in \acceptingset' \}$.
  By the "fundamental lemma of logical relations", for any term $M$ of simple type $A$, we have
  $\semlam{M}{Q} \in \acceptingset$
  if and only if
  $\semlam{M}{Q'} \in \acceptingset'$ as $\semrel{A}{f}$ is a partial function.
  We conclude that $L$ is equal to
  $\semlaminv{Q}(\acceptingset)$,
  so that $L$ is also recognized by~$Q$.
\end{proof}
In order to prove that the union $\reg{A}$ of the Boolean algebras
$\regset{Q}{A}$ is again a Boolean algebra, we will apply the
following general principle from universal algebra in the case
where $\mathbf{V}$ is the variety of Boolean algebras, see for example
\cite[Rem.~3.4.4(iii) on p. 136]{adamek-rosicky}.
\begin{prop}\label{prop/forget-creates-dir-colim}
  For any finitary variety of algebras $\mathbf{V}$, the forgetful functor
  $\mathbf{V} \to \mathbf{Set}$ creates directed colimits.
\end{prop}

We are now ready to prove our first main result, "Theorem~A" of the introduction.

\medbreak
\noindent
\textbf{"Theorem~A".}
\emph{The diagram of Boolean algebras $\regset{(-)}{A} : \FinPSurj^{\mathrm{op}} \to \BA$, i.e.}
\begin{center}\begin{tikzcd}[column sep = 1.2em]
    \Big(\,\, %\label{equation/regA-diagram}
    \regset{Q'}{A}  \arrow[hook,rr] && \regset{Q}{A}
    \Big)_{f:Q\twoheadrightarrow Q'\in\FinPSurj}
    \quad,
  \end{tikzcd}\end{center}
\emph{is directed, and its colimit
  in $\BA$ coincides with the Boolean
  algebra~$\reg{A}$ of regular languages
  of higher-order type~$A$.}
  
\begin{proof}
We first show that the diagram of inclusions of Boolean algebras
is directed.
Indeed, for any finite sets $Q_1$ and $Q_2$, we have, for $i = 1, 2$, the "partial surjection" $f_i \colon Q_1 + Q_2 \twoheadrightarrow Q_i$
defined by
$q \, f_i \, q'$ if and only if $q \in Q_i$ and $q = q'$.
Thus, Lemma~\ref{lemma/partial-surj-inclusion} gives
that $\regset{Q_i}{A} \subseteq \regset{Q_1+Q_2}{A}$.
Now, by Proposition~\ref{prop/forget-creates-dir-colim}, applied in the
case $\mathbf{V} = \mathbf{BA}$, the union $\reg{A}$ of the sets
in the diagram is again a Boolean algebra, and it is the colimit
of the diagram in $\mathbf{BA}$.
\end{proof}

\medbreak

We end this section by showing explicitly how we recover in this
context the result of \cite[Thm.~8]{salvati} that $\reg{A}$
is closed under binary intersection.
\begin{prop}\label{prop/regBA}
  For any simple type $A$, the set of "regular languages" $\reg{A} \subseteq \wp(\LambdaTerms{A})$
  is closed under binary intersection.
\end{prop}
\begin{proof}
  Suppose that $L_1 \in \regset{Q_1}{A}$ and
  $L_2 \in \regset{Q_2}{A}$. By the argument given in the proof of "Theorem~A",
  both $L_1$ and $L_2$ are in $\regset{Q_1+Q_2}{A}$ which is a Boolean algebra,
  so their intersection $L_1 \cap L_2$ is also in $\regset{Q_1+Q_2}{A}$.
\end{proof}

\section{The space of profinite \texorpdfstring{$\lambda$}{λ}-terms}\label{section/profinite}
The aim of this section is to define "profinite $\lambda$-terms"
of an arbitrary simple type $A$
as special "parametric families of semantic elements",
and to show that they form a Stone space dual to
the Boolean algebra~$\reg{A}$.
% ; for a recap of
% the basics of "Stone duality" that we will use below, see
% Appendix~\ref{appendix/stone}.

Throughout this section, we fix a simple type $A$.
We saw in the previous section that~$\reg{A}$
is a Boolean algebra which is the colimit of a directed diagram
of inclusions between the Boolean
algebras~$\regset{Q}{A}$. As~$\regset{Q}{A}$ is
finite for every $Q$, it is isomorphic to $\wp(X_Q(A))$,
where $X_Q(A)$ is the set of atoms of $\regset{Q}{A}$.
Applying discrete Stone duality to the directed diagram of inclusions
of finite Boolean algebras, we thus obtain a codirected diagram of
maps $X_Q(A) \twoheadrightarrow X_{Q'}(A)$, still indexed
by "partial surjections" $Q \twoheadrightarrow Q'$.
We now first give a more concrete description of that diagram.
\begin{prop}\label{prop/atomsofreg}
  For every finite set $Q$, the set of atoms $X_Q(A)$ of $\regset{Q}{A}$
  is in a bijection with the set $\definable{A}{Q}$ of definable
  elements of simple type $A$, given by the function
  \[
    \begin{matrix}
      \definable{A}{Q} & \longrightarrow & X_Q(A)                     \\
      q                & \longmapsto     & \semlaminv{Q}(\{q\})\ .
    \end{matrix}
  \]
\end{prop}
\begin{proof}
  The Boolean algebra $\regset{Q}{A}$ is, by definition, the
  image
  of the Boolean algebra homomorphism
  $\semlaminv{Q} :
    \wp(\semset{A}{Q})
    \to
    \wp(\LambdaTerms{A})$
  Thus, $\regset{Q}{A}$ arises as the following epi-mono factorization of Boolean algebras
  \begin{center}
    \begin{tikzcd}[column sep = 2.5em, row sep = .5em]
      \wp(\semset{A}{Q})
      \arrow[rr,"\semlaminv{Q}"]
      \arrow[rd,two heads,"\semlaminv{Q}"{swap}]
      & & \wp(\LambdaTerms{A})
      \\
      & \regset{Q}{A} \arrow[ru,hook,"p"{swap}]
    \end{tikzcd}
  \end{center}
  Applying the discrete duality functor
  $\mathbf{At} \colon \mathbf{CABA} \to \Set$ to this diagram,
  we get the dual epi-mono factorization of sets
  \begin{center}
    \begin{tikzcd}[column sep = 2.5em, row sep = .5em]
      \semset{A}{Q}
      & & \LambdaTerms{A} \arrow[ld,two heads,""] \arrow[ll,swap,"\semlam{{-}}{Q}"]
      \\
      & X_Q(A) \arrow[lu,hook,""]
    \end{tikzcd}
  \end{center}
  Since $\definable{A}{Q}$ is by definition the image of
  $\semlam{{-}}{Q}$ in $\semset{A}{Q}$, the result follows by the
  uniqueness up to isomorphism of epi-mono factorizations in $\mathbf{CABA}$.
\end{proof}

We now show that "logical relations" induced by "partial surjections", when restricted to "definable elements", all yield the same total function.
% In order to reformulate this property about "partial surjections", we introduce the next proposition.
% which results from Propositions~\ref{prop/fund-lemma}~and~\ref{prop/partial-surjections}.
%
\begin{prop}\label{prop/partial-surj-same}
  For any "partial surjection" $f \colon Q \twoheadrightarrow Q'$ and
  for any simple type $A$,
  the set $\definable{A}{Q} \subseteq \semset{A}{Q}$ of
  definable elements is contained in the domain of $\semrel{A}{f}$,
  and the restriction of $\semrel{A}{f}$ to
  $\definable{A}{Q}$ is the
  unique function $p_{Q,Q'}$ that makes the following diagram commute:
  \begin{center}
    \begin{tikzcd}[column sep = 3em, row sep = 1.5em]
      & \LambdaTerms{A}
      \arrow[dr,two heads,"{\definable{-}{Q'}}"]
      \arrow[dl,two heads,swap,"{\definable{-}{Q}}"]
      \\
      \definable{A}{Q} \arrow[rr,two heads,"p_{Q,Q'}"]
      & &
      \definable{A}{Q'}
    \end{tikzcd}
  \end{center}
\end{prop}
\begin{proof}
  By the "fundamental lemma of logical relations",
  for any term $M$ of simple type $A$, we have
  %inline
  $\semlam{M}{Q} \quad \semrel{A}{f} \quad \semlam{M}{Q'}$, 
  so that any definable element is in the domain of $\semrel{A}{f}$.
  By Proposition~\ref{prop/partial-surjections},
  $\semrel{A}{f}$ is in particular a partial function, so that it
  makes the diagram commute. For the uniqueness, simply note that
  any $q \in \definable{A}{Q}$ can by definition be written
  as $\semlam{M}{Q}$ for some term $M$ of simple type $A$,
  and must therefore be sent
  to $\semlam{M}{Q'}$ by any function making the diagram commute.
\end{proof}

Note that Proposition~\ref{prop/partial-surj-same}
in particular implies that, if
$f, g \colon Q \rightrightarrows Q'$
are two "partial surjections", then,
while their semantic interpretations
$\semrel{A}{f}, \semrel{A}{g} \colon \semset{A}{Q} \rightrightarrows \semset{A}{Q'}$
are in general distinct, their restrictions to the set $\definable{A}{Q}$ of definable
elements
must both be equal to the function $p_{Q,Q'}$.

We are now ready to define "profinite $\lambda$-terms" of a given simple type $A$.
\begin{definition}
  \label{def/prolambda}
  \AP Let $A$ be any simple type. We define the set of ""profinite $\lambda$-terms"" as the limit $\intro*\ProLambdaTerms{A}$ in~$\Set$ of the diagram
  \vspace{-.5em}
  \[
    \begin{tikzcd}[column sep = 1.2em]
      \Big(\,\,
      \definable{A}{f}
      \hspace{.5em}
      :
      \hspace{.5em}
      \definable{A}{Q} \arrow[two heads,rr] && \definable{A}{Q'}\Big)_{f:Q\twoheadrightarrow Q'\in\FinPSurj}
    \end{tikzcd}
    \ .
  \]
\end{definition}
It follows from the way that one calculates limits in $\Set$ that,
concretely, a "profinite $\lambda$-term" in $\ProLambdaTerms{A}$ is a family
$\theta$ of definable elements
$\theta_Q \in \definable{A}{Q}$
where $Q$ ranges over all finite sets
such that
\begin{equation}\label{equation/pro-para-condition1}
\text{for every partial surjection }
  f : Q \twoheadrightarrow Q'\ ,
  \qquad
  \text{we have }
  \definable{A}{f}(\theta_Q)\ =\ \theta_{Q'}
  \ .
\end{equation}

By Proposition~\ref{prop/partial-surj-same},
the condition (\ref{equation/pro-para-condition1}) on the family
$\theta$
is equivalent to the condition that, for any term~$M$ of simple type $A$ and any finite sets $Q$ and $Q'$,
\begin{equation}\label{equation/pro-para-condition2}
  \text{if } \theta_Q = \semlam{M}{Q} \text{ and } |Q| \geq |Q'| \text{ then } \theta_{Q'} = \semlam{M}{Q'} \ .
\end{equation}

We conclude this section
by equipping the set $\ProLambdaTerms{A}$ with
a natural topology, and showing that
this topology turns $\ProLambdaTerms{A}$
into the Stone dual space
of the Boolean algebra $\reg{A}$.
The easiest way to define the topology of $\ProLambdaTerms{A}$ is to say
that it is the subspace topology inherited
from the inclusion
\begin{center}
  \begin{tikzcd}[column sep = 1em]
    \ProLambdaTerms{A}\arrow[hook,rr]
    &&
    {\prod_Q \definable{A}{Q}}
  \end{tikzcd}
\end{center}
into the product space $\prod_Q \definable{A}{Q}$
computed in the category~$\Top$ of topological spaces,
where each component $\definable{A}{Q}$
is considered as a topological space
equipped with the discrete topology.
More concretely, for any finite set $Q$ and $q \in \definable{A}{Q}$,
let us write $U_{Q,q}$ for
the set of "profinite $\lambda$-terms" that take value $q$ at $Q$,
that is,
\[ U_{Q, q}
  \ :=\
  \left\{\quad \theta \in \ProLambdaTerms{A} \ \mid \ \theta_Q = q \quad\right\}
  \ .\]
The topology on $\ProLambdaTerms{A}$ is now defined by taking
the collection of sets $U_{Q,q}$ as a basis, where
$Q$ ranges over all finite sets and $q$ ranges over
all the elements of $Q$.
The following result is proved via an argument
similar to the one given in \cite[Sec.~4.2]{Geh2016} for
profinite algebras.
\begin{prop}\label{prop/profinite-dual-reg}
  The space $\ProLambdaTerms{A}$ is the Stone dual space of
  the Boolean algebra $\reg{A}$. In particular, $\reg{A}$ is
  isomorphic to the Boolean algebra of clopen sets of $\ProLambdaTerms{A}$.
\end{prop}
\begin{proof}
  % As recalled in Appendix~\ref{appendix/stone},
  Stone duality
  arises from the dual equivalence between $\FinSet$ and $\mathbf{FinBA}$
  by taking the projective and inductive completions, respectively.
  Therefore, we have in particular that
  $\ProLambdaTerms{A}$, which is defined as the codirected limit of the
  diagram of finite discrete spaces~$\definable{A}{Q}$ in $\mathbf{Top}$,
  is the dual space of the directed colimit of the diagram
  of finite Boolean algebras $\regset{Q}{A}$, which is the Boolean algebra
  $\reg{A}$ by "Theorem~A".
  The second statement  now follows because
  any Boolean algebra is isomorphic to the collection of clopen sets
  of its dual space.
\end{proof}

\section{Profinite \texorpdfstring{$\lambda$}{λ}-terms and parametricity}
\label{section/profiniteness-and-parametricity}

\AP
Let $A$ be any simple type. A ""parametric family"" is a family of points $\theta_Q \in \semset{A}{Q}$, where $Q$ ranges over all finite sets, such that for any relation $R \subseteq Q \times Q'$, we have
$\theta_Q\ \semrel{A}{R}\ \theta_{Q'}$.

Every "parametric family" $\theta$ is in particular parametric with respect to "partial surjections". Therefore, a "parametric family" whose components are "definable elements" is a "profinite $\lambda$-term". We now show that the converse holds.

\medbreak
\noindent\textbf{"Theorem~B".}
\emph{
  A "profinite $\lambda$-term" $\theta \in \ProLambdaTerms{A}$
  of simple type~$A$
  may be equivalently defined as a "parametric family" of 
  "definable elements" $\theta_Q\in\definable{A}{Q}$.}

\begin{proof}
  Let $\theta$ be a "profinite $\lambda$-term", viewed as a family of
  definable elemens which is parametric with respect
  to every partial surjection, or equivalently, satisfying
  condition (\ref{equation/pro-para-condition2}).
  Let $Q_1$ and $Q_2$ be any two finite sets and let $R \subseteq Q_1 \times Q_2$ be any relation.
  Pick any finite set $Q$ of cardinality $\max(|Q_1|,|Q_2|)$.
  Since $\theta_{Q}$ is
  in particular definable, pick a $\lambda$-term $M$ in $\LambdaTerms{A}$ such that
  %inline
  $\theta_{Q}$ is $\semlam{M}{Q}$.
  Since $|Q| \geq |Q_i|$ for $i = 1,2$,
  by (\ref{equation/pro-para-condition2}) we now also have
  %inline
  $\theta_{Q_i}$ is equal to
  $\semlam{M}{Q_i}$.
  By
  % Proposition~\ref{prop/fund-lemma}
  the "fundamental lemma of logical relations",
  we obtain that
  %inline
  $\semlam{M}{Q_1}\ \semrel{A}{R}\ \semlam{M}{Q_2}$
  which proves that $\theta$ is a "parametric family".
\end{proof}

\section{The cartesian closed category of profinite~\texorpdfstring{$\lambda$}{λ}-terms}\label{section/profinite-ccc}

\AP
We now show that "profinite $\lambda$-terms" assemble into a cartesian closed category~$\ProLam$ which thus provides an interpretation of the simply typed $\lambda$-calculus.
In order to construct the category~$\ProLam$,
we find it convenient to use a general construction
introduced by Jacq and Melliès~\cite{jacq-mellies}
in a more general monoidal and 2-categorical setting.
Suppose given a cartesian closed category $\C$ and a functor
\[\mathscr{P}
  \quad:\quad
  \C
  \ \longrightarrow\
  \Set
\]
which is ""cartesian product preserving"" in the sense that the canonical functions
\begin{align*}
  \langle \mathscr{P}(\pi_1), \mathscr{P}(\pi_2)\rangle
  \quad & :\quad
  \mathscr{P}(A \times B)
  \ \longrightarrow\
  \mathscr{P}(A) \times \mathscr{P}(B)
  \\
  !_{\mathscr{P}(1)}
  \quad & :\quad
  \mathscr{P}(1)
  \ \longrightarrow\
  1
\end{align*}
are bijections for all objects $A$ and $B$
of the category~$\C$.
We denote by
\begin{align*}
  m_{A,B}
  \quad & :\quad
  \mathscr{P}(A) \times \mathscr{P}(B)
  \ \longrightarrow\
  \mathscr{P}(A \times B)
  \\
  m_{1}
  \quad & :\quad
  1
  \ \longrightarrow\
  \mathscr{P}(1)
\end{align*}
the inverse functions.
In that situation, one defines the category $\C[\mathscr{P}]$ whose objects are the objects of~$\C$ and whose hom-sets are defined as follows:
\[\C[\mathscr{P}](A,B) \quad := \quad \mathscr{P}(A \To B)\]
using the internal hom-object $A\Rightarrow B$ of the cartesian closed category~$\C$. Equivalently, $\C[\mathscr{P}]$ is the result of seeing the cartesian closed category~$\C$ as enriched in itself, and then changing the base along $\mathscr{P}$.
One establishes that

\begin{prop}\label{proposition/ccc-from-P}
  The category~$\C[\mathscr{P}]$ is cartesian closed
  and comes equipped with a cartesian closed
  identity-on-object functor
  \begin{center}
    \begin{tikzcd}[column sep=1em]
      \inclusion{\C,\mathscr{P}}
      \quad : \quad
      \C\arrow[rr] && {\C[\mathscr{P}]}
    \end{tikzcd}
  \end{center}
  which strictly preserves the cartesian product as well as the internal hom.
\end{prop}
% The interested reader will find the proof of Proposition~\ref{proposition/ccc-from-P} in Appendix~\ref{appendix/proofs-profinite-ccc}.
%
Now, in order to obtain the category~$\ProLam$ of "profinite $\lambda$-terms" using this categorical construction, we start by recalling the definition of the cartesian closed category~$\Lam$ freely generated by the terminal category.

\begin{definition}
  \AP The category~$\intro*\Lam$ has as objects the simple types of the $\lambda$-calculus and its hom-sets are defined as
  \[\Lam(A, B)
    \ :=\
    \LambdaTerms{A \To B}
    \
  \]
  for all pairs $A$ and $B$ of simple types.
\end{definition}
At this stage, we are ready to consider the functor
%inline
$\mathscr{P} : \Lam \longrightarrow \Set$
which transports every simple type~$A$
to the set of "profinite $\lambda$-terms"
\begin{equation}\label{equation/functor-P-for-ProLam}
  \mathscr{P}(A) \quad = \quad \ProLambdaTerms{A}
\end{equation}
and every $\lambda$-term $M$ of simple type $A \To B$ to the set-theoretic function sending a "profinite $\lambda$-term" $\theta$ of simple type $A$ on the family $(\semlam{M}{Q}(\theta_Q))$ which can be shown to be a "profinite $\lambda$-term" of simple type $B$ using
% Proposition~\ref{prop/fund-lemma}
the "fundamental lemma of logical relations".
It is interesting to observe that the functor~$\mathscr{P}$ is "cartesian product preserving"
and that we have canonical bijections
\[
  \mathscr{P}(A\times B) \,\cong\, \mathscr{P}(A)\times\mathscr{P}(B)
  \quad\quad\quad
  \mathscr{P}(1) \,\cong \,1
\]
for every pair of simple types~$A$ and $B$.
By applying the construction, we obtain a cartesian closed category
\[\intro*\ProLam \  := \  \Lam[\mathscr{P}]\]
whose objects are the simple types of the $\lambda$-calculus and whose hom-sets are defined as follows:
\[\ProLam(A, B)
  \ :=\
  \ProLambdaTerms{A \To B}
  \ .
\]

\begin{remark}
  Note that the functors $\mathscr{P}$ are chosen to be valued in $\Set$, but we could choose any cartesian category $\cat{S}$ as long as $\mathscr{P}$ still is "cartesian product preserving" relatively to the cartesian structure of $\cat{S}$. The construction will then yield a cartesian closed category $\C[\mathscr{P}]$ enriched over $\cat{S}$. As a matter of fact, the functor $\mathscr{P}:\Lam\to\Set$
  used in \eqref{equation/functor-P-for-ProLam}
  to construct $\ProLam=\Lam[\mathscr{P}]$ happens to
  factor through the category $\Stone$ of Stone spaces,
  in the following way:
  \begin{center}
    \begin{tikzcd}
      \Lam\arrow[rr,"{\ProLambdaTerms{-}}"]
      &&
      \Stone
      \arrow[rr,"{\text{forget}}"]
      &&
      \Set
    \end{tikzcd}
  \end{center}
  This shows that the cartesian closed category $\ProLam$ of "profinite $\lambda$-terms" may be also considered as enriched over the category~$\Stone$ of Stone spaces.
\end{remark}

\section{A faithful embedding from \texorpdfstring{$\lambda$}{λ}-terms to profinite \texorpdfstring{$\lambda$}{λ}-terms}
\label{section/faithful-embedding}
By construction, the category~$\ProLam$ comes equipped with a cartesian closed identity-on-object functor
\begin{equation}\label{equation/from-lam-to-profinite}
  \begin{tikzcd}[column sep=1em]
    \inclusion{}{}
    \quad : \quad
    \Lam\arrow[rr] && {\ProLam}
  \end{tikzcd}
\end{equation}
which may also be derived from the fact
that $\Lam$ is the free cartesian closed category.
We now establish that
\begin{prop}\label{proposition/faithful}
  The functor $\inclusion{}{}$ is faithful.
\end{prop}

Towards proving Proposition~\ref{proposition/faithful},
we first claim that the category $\ProLam$
can be obtained as the limit of a codirected diagram
of cartesian closed categories, described in the following way.
\AP
Given a finite set~$Q$ and a simple type~$A$, consider the equivalence relation
\[\congruence{A}{Q} \quad \subseteq \quad \LambdaTerms{A}\times\LambdaTerms{A}\]
on the set of simply typed $\lambda$-terms of type~$A$ modulo $\beta\eta$-conversion, defined as:
\[
  M\intro*\congruence{A}{Q} N \ \iffdef \ \semset{M}{Q}=\semset{N}{Q} \ .
\]
When we fix the finite set~$Q$,
the family $\congruence{}{Q}$
of equivalence relations $\congruence{A}{Q}$
parametrized by simple types~$A$
defines a congruence relation
on the category~$\Lam$,
in the expected sense that
\begin{center}
  if
  $f\congruence{A\Rightarrow B}{Q}f'$
  \, {and} \,
  $g\congruence{B\Rightarrow C}{Q}g'$,
  \, then \,
  $g\circ f \congruence{A\Rightarrow C}{Q} g'\circ f'$
\end{center}
for any tuple of morphisms $f,f',g,g'$ of the form:
\begin{center}
  \begin{tikzcd}[column sep=2em]
    A \arrow[rr,yshift=.4em,"f"]\arrow[rr,yshift=-.4em,"{f'}"{swap}]
    &&
    B\arrow[rr,yshift=.4em,"g"]\arrow[rr,yshift=-.4em,"{g'}"{swap}]
    &&
    C
  \end{tikzcd}
  \ .
\end{center}
\AP
From this, it follows that we can define the category
\[\intro*\Lamof{Q} \ := \ \Lam\,/\congruence{}{Q}\]
obtained by considering the morphisms of the free cartesian closed category $\Lam$ modulo the congruence relation $\congruence{}{Q}$ in the expected sense that
\begin{equation}\label{equation/hom-set-quotiented}
  \Lamof{Q}(A,B) \, = \, \LambdaTerms{A\Rightarrow B} \, / \congruence{A\Rightarrow B}{Q}\ .
\end{equation}
We then establish
% (see Appendix~\ref{appendix/proofs-profinite-ccc} for the proof)
that
\begin{prop}\label{proposition/diagram-of-lamQ}
  For every finite set~$Q$, the category $\Lamof{Q}$ is cartesian closed and comes equipped with a cartesian closed identity-on-object functor
  \begin{center}
    \begin{tikzcd}[column sep=1em]
      \pi_{Q}
      \quad : \quad
      \Lam\arrow[rr] && {\Lamof{Q}}
      \ .
    \end{tikzcd}
  \end{center}
  Moreover, every partial surjection $f:Q\twoheadrightarrow Q'$ in the category $\FinPSurj$
  induces a cartesian closed identity-on-object functor
  \begin{center}
    \begin{tikzcd}[column sep=1em]
      \Lamof{f}
      \quad : \quad
      {\Lamof{Q}}\arrow[rr] && {\Lamof{Q'}}
    \end{tikzcd}
  \end{center}
  making the diagram of cartesian closed functors commute:
  \begin{center}
    \begin{tikzcd}[column sep=1em, row sep = .2em]
      &&&&
      \Lamof{Q}\arrow[dddd,"{\Lamof{f}}"]
      \\
      \\
      \Lam\arrow[rrrruu,bend left = 15, "{\pi_{Q}}"]
      \arrow[rrrrdd,bend right = 15, "{\pi_{Q'}}"{swap}]
      \\
      \\
      &&&& \Lamof{Q'}
    \end{tikzcd}
  \end{center}
\end{prop}
\noindent
From this observation, it is not too difficult
to show that
\begin{prop}
  The category $\ProLam$ is the codirected limit of the diagram of cartesian closed categories $\Lamof{Q}$ indexed by finite sets and partial surjections.
  The projection functor
  \begin{center}
    \begin{tikzcd}[column sep=1em]
      \pi_{Q}
      \quad : \quad
      \ProLam\arrow[rr] && {\Lamof{Q}}
    \end{tikzcd}
  \end{center}
  is defined by transporting every morphism $\theta:A\to B$ defined as a family~$\theta$ of definable elements in
  \begin{center}
    $\ProLam(A,B) \quad = \quad \ProLambdaTerms{A\Rightarrow B}$
  \end{center}
  to the instance $\theta_Q$ in
  \[\Lamof{Q}(A,B) \, = \,
    \LambdaTerms{A\Rightarrow B} \, / \congruence{A\Rightarrow B}{Q}
    \, \cong \, \definable{A\Rightarrow B}{Q}
    \ .
  \]
\end{prop}
\noindent
One also establishes that the canonical functor~\eqref{equation/from-lam-to-profinite}
is also characterized by the fact that it is the unique cartesian closed functor making the diagram below commute:
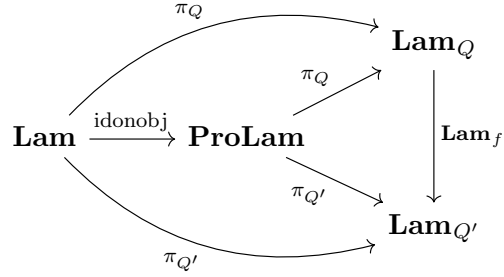
\begin{center}
  \begin{tikzcd}[column sep=1em, row sep = .8em]
    &&&
    &&
    \Lamof{Q}\arrow[dddd,"{\Lamof{f}}"]
    \\
    \\
    \Lam\arrow[rrr,"{\inclusion{}{}}"]\arrow[rrrrruu,bend left = 30,"{\pi_{Q}}"]\arrow[rrrrrdd,bend right = 30,"{\pi_{Q'}}"{swap}]
    &&&
    \ProLam\arrow[rruu,"{\pi_{Q}}"]
    \arrow[rrdd,"{\pi_{Q'}}"{swap}]
    \\
    \\
    &&&
    && \Lamof{Q'}
  \end{tikzcd}
\end{center}
This observation provides us with a clean proof that the functor $\inclusion{}{}$ is faithful.
Indeed, by Statman's finite completeness theorem~\cite{statman82},
for every pair of morphisms $f, g : A \rightrightarrows B$
in the category~$\Lam$ which is (by definition) a pair of $\lambda$-terms $M$ and $N$ modulo $\beta\eta$-conversion, either $M$ and $N$ are equal modulo $\beta\eta$-conversion or there exists a finite set~$Q$ such that the interpretations $\semlam{M}{Q}=\pi_Q(M)$ and $\semlam{N}{Q}=\pi_Q(N)$ are different.
In particular, $\inclusion{}{}(f)$ and $\inclusion{}{}(g)$
differ in the second case.
This establishes that the canonical functor $\inclusion{}{}:\Lam\to\ProLam$ is faithful,
as claimed in Proposition~\ref{proposition/faithful}.

\section{Profinite \texorpdfstring{$\lambda$}{λ}-terms and profinite words}
\label{section/proterm-proword}

The higher-order language theory on simply typed $\lambda$-terms is designed to extend the traditional language theory on words on a given finite alphabet $\Sigma$.
The idea is that a finite word on the alphabet~$\Sigma$ is the same thing as a $\lambda$-term of simple type $\Churchs$ modulo $\beta\eta$-conversion.
In particular, we recall below a folklore result which states that the Boolean algebra $\reg{\Churchs}$ of regular higher-order languages on $\Churchs$ coincides with the Boolean algebra $\operatorname{Reg}\langle\Sigma\rangle$
of regular languages on the finite alphabet~$\Sigma$.
%
%We will use that correspondence
%in order to establish to show "Theorem~C".

\begin{proposition}
  \label{prop/reg-reg-same}
For every finite alphabet~$\Sigma$, 
one has an isomorphism of Boolean algebra
\[
\reg{\Churchs}
\quad \cong \quad
\operatorname{Reg}\langle\Sigma\rangle
\]
given by the Church encoding.
\end{proposition}

\begin{proof}
The Church encoding provides a one-to-one correspondence between subsets~$L\subseteq \Sigma^{\ast}$ of words over the alphabet $\Sigma$ and subsets~$\mathcal{L}\subseteq\LambdaTerms{\Churchs}$ of $\lambda$-terms of simple type $\Churchs$ closed modulo $\beta\eta$-conversion.
We show that a subset~$L\subseteq \Sigma^{\ast}$ is regular if and only if the associated subset~$\mathcal{L}\subseteq\LambdaTerms{\Churchs}$ is an element of $\reg{\Churchs}$.

In one direction, suppose that $L\subseteq\Sigma^{\ast}$ is a language of words recognized by a DFA $\automaton = (Q, \delta, q_0, \acceptingset)$. We recall from the introduction that the associated set~$\mathcal{L}\subseteq\LambdaTerms{\Churchs}$ of $\lambda$-terms
is the inverse image by the 
semantic bracket
\[
\begin{tikzcd}[column sep = 1em]
\semlam{{-}}{Q} \quad : \quad \LambdaTerms{\Churchs} \arrow[rr] && \semset{\Churchs}{Q} 
\end{tikzcd}
\]
of the set of functionals in $\semset{\Churchs}{Q}$ defined as follows
\[\evalfun{(\delta,q_0)}^{-1}(\acceptingset)
\ =\ 
\{F \in \semset{\Churchs}{Q} \mid F(\delta_{a_1}, \dots, \delta_{a_n})(q_0) \in \acceptingset\}
\ .
\]
By definition, $\mathcal{L}\subseteq\LambdaTerms{\Churchs}$
is thus an element of $\regset{Q}{\Churchs}$
and thus an element of $\reg{\Churchs}$.

Conversely, by definition of $\reg{\Churchs}$, it is sufficient to establish, for every finite set~$Q$, that every subset $\mathcal{L}\in\regset{Q}{\Churchs}$
has its corresponding subset~$L\subseteq \Sigma^{\ast}$ a regular language.
By definition of $\regset{Q}{\Churchs}$, we may suppose without loss of generality that $\mathcal{L}\in\regset{Q}{\Churchs}$ is of the form
\[
\mathcal{L}
\ =\ 
\semlaminv{Q}(\{F\})
\ =\ 
\{M\in\LambdaTerms{\Churchs} \mid \semlam{M}{Q}=F\}
\]
where $F$ is a functional in $\semset{\Churchs}{Q}$.
The corresponding set~$L\subseteq\Sigma^{\ast}$ is the finite intersection of all the regular languages recognized by the DFAs of the form $\A = (Q, \delta, q_0, \{q_f\})$ where the unique final state
$q_f$ is equal to $q_f=F(\delta_{a_1}, \dots, \delta_{a_n})(q_0)$.
As a finite intersection of regular languages, the set~$L\subseteq\Sigma^{\ast}$ is itself regular.
\end{proof}

We use this result in order to establish our "Theorem~C".

\medbreak
\noindent
\textbf{"Theorem~C".}
\emph{For every finite alphabet $\Sigma$, there is a homeomorphism
\[
\ProLambdaTerms{\Churchs} \quad \cong \quad \widehat{\Sigma^*}
\]
between the space $\ProLambdaTerms{\Churchs}$ of "profinite $\lambda$-terms" of type $\Churchs$ and the space $\widehat{\Sigma^*}$ of profinite words.}

\begin{proof}
By Proposition \ref{prop/profinite-dual-reg},
the space $\ProLambdaTerms{\Churchs}$ 
is the Stone dual of the Boolean algebra
$\reg{\Churchs}$
which is isomorphic to
the Boolean algebra $\operatorname{Reg}\langle\Sigma\rangle$ by Proposition~\ref{prop/reg-reg-same}.
From this follows that the space $\ProLambdaTerms{\Churchs}$
is homeomorphic to the Stone dual
of $\operatorname{Reg}\langle\Sigma\rangle$
which coincides with 
the space $\widehat{\Sigma^*}$ of
profinite words by an important result
of Stone duality, see~\cite{pin2009}.
\end{proof}

One main benefit of extending finite words into profinite words is that a new class of implicit operations become available \cite{almeidabook}.
In particular, there exists an ""idempotent power operator"" $u\mapsto u^{\omega}$ which turns every profinite word $u$ into another profinite word noted $u^{\omega}$,
and defines a continuous function
\begin{equation}\label{equation/idempotent-power-on-words}
u \longmapsto u^{\omega}
\quad : \quad
\widehat{\Sigma^*}
\ \longrightarrow\ 
\widehat{\Sigma^*}
\end{equation}
 see for example~\cite[Prop.~2.5]{pin2009}.
The construction is based on the observation that for any element $x$ of a finite monoid $M$, there exists a unique power $x^n$ of $x$, for $n \ge 1$, which is idempotent.
This unique power is obtained when $n$ is the factorial of the cardinality of $M$, and is also usually written $x^\omega$.
The continuous function~\eqref{equation/idempotent-power-on-words} is obtained by taking the profinite limit of this operation on monoids.
We show the construction generalizes from profinite words to "profinite $\lambda$-terms" at every type~$A$.

\begin{proposition}
 For every simple type $A$, there exists a "profinite $\lambda$-term"
 \begin{equation}\label{equation/idempotent-power-on-terms}
\Omega_A \quad : \quad (A\To A) \To A \To A
\end{equation}
which, given any $M \in \ProLambdaTerms{A \To A}$, satisfies the idempotency equation
\[
(\Omega_A\,M) \circ (\Omega_A\,M) \quad = \quad \Omega_A\,M
\]
between "profinite $\lambda$-terms",
where $g\circ f$ is notation for $\lambda x. f\,(g\,x)$ where $f$ and $g$ are "profinite $\lambda$-terms".
\end{proposition}

%\begin{proof}
%    By definition, $\Omega_A$ is a family of "definable elements". We use the characterization of Equation~\ref{equation/pro-para-condition2} to show that it is natural with respect to partial surjections. Let $Q$ and $Q'$ be any two finite sets such that $|Q|$ is greater than $|Q'|$. Let $n$ and $n'$ be the respective cardinalities of $\semset{A}{Q}$ and $\semset{A}{Q'}$. By definition of $\Omega_A$, we have $\Omega_{A, Q} = \semlam{\underline{n!}}{Q}$ so let us show that $\Omega_{A, Q'} = \semlam{\underline{n!}}{Q}$. To this end, we consider any $f : Q \to Q$ and want to show that $\semlam{\underline{n'!}}{Q}(f)$ and $\semlam{\underline{n!}}{Q}(f)$ are equal. As $f^{n'!}$ is the idempotent power of $f$ and $n'!$ divides $n!$, $f^{n!}$ is equal to $f^{n'!}$, which concludes the proof.
%\end{proof}

We have seen in "Theorem~C" that one recovers the traditional notion of profinite words on a finite alphabet~$\Sigma$ by considering the "profinite $\lambda$-terms" of type $\Churchs$.
Accordingly, the continuous operation~\eqref{equation/idempotent-power-on-words} can be recovered as the "profinite $\lambda$-term"
\[
\begin{tikzcd}[column sep = 1em]
\lambda u. \lambda f_1\dots f_n. \Omega_\tyo\,(u\,f_1\,\dots\,f_n)
\quad : \quad 
\Churchs \To \Churchs
\end{tikzcd}
\]
where $\Omega_{\tyo}:(\tyo\To\tyo)\To(\tyo\To\tyo)$ denotes the idempotent power operator~$\Omega_A$ at type $A=\tyo$.
Note that we use the compositional calculus provided by the cartesian closed category $\ProLam$ in order to see the expression $\lambda u. \lambda f_1\dots f_n.\,\Omega_\tyo\,(u\,f_1\,\dots\,f_n)$ as a "profinite $\lambda$-term".

\section{Conclusion}
\label{section/conclusion}
In this paper, we introduce the notion of \emph{"profinite $\lambda$-term"} of a given simple type which we define in a clean and principled way by establishing in "Theorem~A" and "Theorem~B" that the definitions based on duality theory and on parametricity coincide.
We also establish in "Theorem~C" that the Church encoding of finite words as $\lambda$-terms extends to profinite words, in the sense that the usual notion of profinite word on a finite alphabet~$\Sigma$ coincides with the notion of "profinite $\lambda$-term" on the type $\Churchs$ encoding the alphabet~$\Sigma$.
We also construct a cartesian closed category~$\ProLam$ of "profinite $\lambda$-terms",
and construct a cartesian closed functor
\begin{center}
\begin{tikzcd}[column sep=1em]
\inclusion{} \quad : \quad \Lam \arrow[rr] && \ProLam
\end{tikzcd}
\end{center}
from the cartesian closed category of usual simply typed $\lambda$-terms.
We also show that this embedding functor from simply typed $\lambda$-terms to "profinite $\lambda$-terms" is faithful, using Statman's theorem.
The construction shows that simply typed $\lambda$-terms can be considered as particular "profinite $\lambda$-terms", and that "profinite $\lambda$-terms" can be manipulated in the same compositional way as usual simply typed $\lambda$-terms.
%

%An important observation not mentioned in the paper for lack of space, but which we would like to develop in future work, is that a "parametric family"~$\theta$ of semantic elements of type~$A$ can be understood as an element of the parametric interpretation in $\FinSet$ of the polymorphic type $\forall \tyo.A$ of System~$F$, in the style of Reynolds~\cite{Reynolds83}.
%
%Seen from that point of view, our parametricity theorem says that, at Church type~$A$, every element of the interpretation of $\forall \tyo.A$ in $\FinSet$ is the denotation of a "profinite $\lambda$-term".
%
%Note that we need to consider "profinite $\lambda$-terms" (and not just $\lambda$-terms) in order to get this parametricity theorem identifying elements of the interpretation $\forall \tyo. A$ with a (profinite) notion of $\lambda$-term.

\bibliographystyle{entics}
\bibliography{biblio}

\end{document}